\documentclass[aps,amsfonts,prl,nofootinbib,tightenlines,showpacs]{revtex4}

\usepackage{bm}
\usepackage{epsfig}

\newcommand{\sech}{{\, \rm sech \,}}
\newcommand{\arxiv}[1]{{#1}}
\newcommand{\currtime}{(t)}

\begin{document}

\title{Numerical Simulation of an Electroweak Oscillon}

\author{N.\ Graham}
\email{ngraham@middlebury.edu}
\affiliation{Department of Physics, Middlebury College,
Middlebury, VT  05753} 

\preprint{\rm hep-th/yymmnnn}

\begin{abstract}
Numerical simulations of the bosonic sector of the $SU(2)\times
U(1)$ electroweak Standard Model in $3+1$ dimensions have
demonstrated the existence of an oscillon --- an extremely
long-lived, localized, oscillatory solution to the equations of motion
--- when the Higgs mass is equal to twice the $W^\pm$ boson mass.  
It contains total energy roughly 30\ TeV localized in a region of
radius 0.05\ fm.  A detailed description of these numerical results is
presented.

\end{abstract}

\maketitle

\section{Introduction}

While static, localized soliton solutions to the equations of motion
of nonlinear field theories have been well studied, and are of
interest in many applications \cite{Coleman,Rajaraman}, no known
examples exist in the electroweak Standard Model (although there do
exist extended  electroweak string solutions \cite{zstring,Volkov}).
However, much less is known about the existence of localized solutions
that oscillate in time, known as breathers or oscillons.  (The latter
term was originally introduced to describe similar phenomena in plasma
physics \cite{Stenflo}.)  In some models, such as the
sine-Gordon breather \cite{DHN} and $Q$-ball \cite{ColemanQ}, one can
use conserved charges to prove the existence of exact, periodic
solutions.  But oscillons have also been found in many nonlinear field
theories that do not contain either static solitons or conserved
charges.  These solutions either live indefinitely or for extremely
long times compared to the natural timescales of the system.

For scalar theories in one space dimension, oscillons have
been found to remain periodic to all orders in a perturbative
expansion \cite{DHN} and are never seen to decay in numerical
simulations \cite{Campbell}, but can decay after extremely long times
via nonperturbative effects \cite{Kruskal} or by coupling to an
expanding background \cite{oscex}.  In both $\phi^4$ theory in two
dimensions \cite{2d1,2d2} and the abelian Higgs model in one dimension
\cite{abelianhiggs} and in two dimensions \cite{GleiserU1}, oscillons
have been found that are not observed to decay.  In $\phi^4$ theory in
three dimensions, however, one finds long-lived quasi-periodic
solutions whose lifetime depends sensitively on the initial conditions
\cite{Bogolyubsky,Gleiser,Honda,iball,Forgacs}.  Similar behavior is
present in other scalar theories in three dimensions \cite{Wojtek} and
in higher dimensions \cite{Gleiserd}.  Phenomenologically, small
$Q$-balls were considered as dark matter candidates in
\cite{smallq1,smallq2,Enqvist,Kasuya}, axion oscillons were considered
in \cite{Kolb}, and the effects of oscillons and other aspects of
nonequilibrium dynamics in and after inflation were studied in
\cite{McDonald,Rajantie,Gleiserinflat}.  Oscillons and related
solutions have also been studied in connection with phase transitions
\cite{Gleiserphase}, monopole systems \cite{monopole}, QCD \cite{Hsu},
and gravitational systems \cite{Khlopov}.

Recent work \cite{oscsm} demonstrated numerically the existence of
an oscillon in the bosonic sector of the electroweak Standard Model, when
the mass of the fundamental Higgs is exactly twice that of the $W^\pm$
gauge bosons.  (A similar mass relation also arises in the study of
embedded defects \cite{Lepora}.)  This result was based on previous
work \cite{oscillon}, which found oscillons in spontaneously broken
pure $SU(2)$ Higgs-gauge theory with the same $2:1$ mass ratio.  In
that model, one can consider field configurations restricted to the
spherical ansatz \cite{spherical}, meaning they are assumed to be
invariant under combined rotations in space and isospin, also known as
grand spin rotations.  Within this ansatz, the system can be described
by an effective theory of fields depending only on $r$ and $t$, which
greatly simplifies the numerical analysis.  In \cite{oscsm} this
numerical simulation was extended to a fully three-dimensional spatial
lattice with no assumptions of rotational symmetry, making it possible
to also include the $U(1)$ hypercharge field (which breaks the grand
spin invariance of the spherical ansatz).  The resulting simulation
comprises the full electroweak sector of the Standard Model without
fermions.  Here we extend that analysis and describe
its results in more detail.  We use the same $SU(2)$ gauge coupling
$g$ and Higgs self-coupling $\lambda$ as in the pure $SU(2)$ theory,
meaning that the Higgs mass is twice the mass of the $W^{\pm}$ bosons,
and set the $U(1)$ coupling $g'$ so that the mass of the $Z^0$ boson
matches its observed value.

Ongoing analytic work \cite{twofield} has shed some light on the
$2:1$ mass ratio by using a small amplitude approximation
\cite{DHN,Hsu,smallamp,oscex} to construct oscillons in a simplified
version of the spherical ansatz theory.  In this analysis, one begins
by assuming that each field in the oscillon profile has large width,
so that at large distances it falls like $\exp(-\epsilon m r)$, where
$m$ is its mass.  There, the amplitude is small and the oscillations
obey a linear dispersion relation, which implies $\omega =
m\sqrt{1-\epsilon^2}$.  The linear, dispersive gradient terms in
the equation of motion are then of order $\epsilon^2$.  They must be
balanced by nonlinear terms to obtain a stable solution.  Since the
leading nonlinearity is typically given by a quadratic term in the
equations of motion, this requirement implies that the field
amplitudes must be proportional to $\epsilon$.  In a multiple-field
model, one must also ensure that the terms giving interactions between
different fields are resonant with the dispersive linear terms,
so that their effects are not washed out over many cycles.  As shown in
\cite{twofield}, the $2:1$ mass ratio arises naturally in this analysis:
Since the fields' oscillation frequencies are tied to their masses,
imposing a resonance condition on their frequencies is equivalent to
fixing a particular mass ratio.  Although this analysis has so far
only been carried out in simplified models, we will see below that the
oscillon observed numerically in the full electroweak theory is of
small amplitude and large width, so similar techniques are potentially
applicable in this case as well.

In all known oscillons, each field oscillates with a frequency below
its mass, so that it couples to dispersive linear waves (which have
$\omega = \sqrt{k^2+m^2} > m$) only through nonlinear interactions.
The fields then converge to a configuration in which this decay channel
is also suppressed.  Because the electroweak theory includes the
massless photon field, which can radiate in arbitrarily low
frequencies, one might expect the oscillon to decay rapidly by emitting
electromagnetic radiation, but it does not.  Instead, after initially
shedding some energy in this way, the system settles into a localized
solution that no longer radiates and remains stable for as long as we
can follow it in numerical simulations.  In preliminary work that provided
motivation for the current investigation, similar behavior was observed
both when an additional massless scalar field was coupled to oscillons
in one-dimensional $\phi^4$ theory and when an additional spherically
symmetric massless scalar field was coupled to oscillons in the
spherical ansatz model.  In each case, after shedding some energy into
the massless field, the oscillon arranges itself in a neutral
configuration that no longer couples to the massless field.  This
mechanism may be similar to the suppression of nonlinear coupling to
dispersive waves that is common to all oscillons.

\section{Continuum Theory}

We begin from $SU(2)\times U(1)$ electroweak theory in the continuum,
ignoring fermions, and follow the conventions of \cite{Huang}.  The
Lagrangian density is
\begin{equation}
{\cal L} = -\frac{1}{4} F_{\mu \nu}
F^{\mu \nu} -\frac{1}{4} {\bm F}_{\mu \nu} \cdot {\bm F}^{\mu \nu}
+ (D_\mu \Phi)^\dagger D^\mu \Phi 
- \lambda(|\Phi|^2 - v^2)^2 \,,
\end{equation}
where the boldface vector notation refers to isovectors.  Here $\Phi$
is the Higgs field, a Lorentz scalar carrying $U(1)$
hypercharge $1/2$ and transforming under the fundamental
representation of $SU(2)$.  The metric signature is $+---$.
The $SU(2)$ and $U(1)$ field strengths are
\begin{equation}
\bm{F}_{\mu \nu} = \partial_\mu {\bm W}_\nu - \partial_\nu \bm{W}_\mu
- g{\bm W}_\mu\times {\bm W}_\nu \,, \qquad
F_{\mu \nu} = \partial_\mu B_\nu - \partial_\nu B_\mu \,,
\end{equation}
and the covariant derivatives are given by
\begin{equation}
D_\mu \Phi = \left(\partial_\mu + i \frac{g'}{2} B_\mu
+ i \frac{g}{2} \bm{\tau} \cdot \bm{W}_\mu\right)\Phi \,, \qquad
D^\mu \bm{F}_{\mu \nu} = \partial^\mu \bm{F}_{\mu \nu}
- g \bm{W}^\mu \times \bm{F}_{\mu\nu} \,,
\end{equation}
where $\bm{\tau}$ represents the weak isospin Pauli matrices.
We obtain the equations of motion
\begin{equation}
\partial_\mu F^{\mu\nu} = J^\nu \,, \qquad
D_\mu \bm{F}^{\mu\nu} = \bm{J}^\nu \,, \qquad
D^\mu D_\mu \Phi = 2\lambda(v^2 - |\Phi|^2) \Phi\,,
\end{equation}
where the gauge currents are
\begin{equation}
J_\nu = g' {\, \rm Im \,} (D_\nu \Phi)^\dagger \Phi\,, \qquad
\bm{J}_\nu = g {\, \rm Im \,} (D_\nu \Phi)^\dagger \bm{\tau} \Phi \,.
\end{equation}

We work in the gauge $B_0 = 0$, $\bm{W}_0 = \bm{0}$.  With this choice, the
covariant time derivatives become ordinary derivatives and we can
apply a Hamiltonian formalism.  The energy density is
\begin{eqnarray}
u &=& \frac{1}{2}\sum_{j=x,y,z}\left[
\dot B_j^2 + \bm{\dot W}_j\cdot\bm{\dot W}_j 
+ \sum_{k>j} \left(
F_{kj}^2 + \bm{F}_{kj} \cdot \bm{F}_{kj} \right) \right]
+ |\dot \Phi|^2 + \sum_{j=x,y,z} (D_j \Phi)^\dagger (D_j \Phi)
+ \lambda\left(|\Phi|^2 - v^2\right)^2 \,,
\end{eqnarray}
where dot indicates time derivative.  The integral over space of this
quantity is conserved by the time evolution.  From the equations for
$B_0$ and $\bm{W}_0$, we obtain the Gauss's Law constraints,
\begin{equation}
\sum_{j=x,y,z} \partial_j \dot B_j - J_0 = 0\,, \qquad
\sum_{j=x,y,z} D_j \bm{\dot W}_j - \bm{J}_0 = 0\,,
\end{equation}
where the charge densities are
\begin{equation}
J_0 = g' {\, \rm Im \,} \dot \Phi^\dagger \Phi \,, \qquad
\bm{J}_0 = g {\, \rm Im \,} \dot \Phi^\dagger \bm{\tau}\Phi \,.
\end{equation}
These constraints remain true at all times, at all points in space,
assuming they are obeyed by the initial value data.

Although the numerical calculation will be done using the underlying
gauge fields ${\bm W}_\mu$ and $B_\mu$, because of spontaneous
symmetry breaking the physical content of the theory is better
described by the fields of definite mass and electric charge
\begin{eqnarray}
W^\pm_\mu &=& \frac{1}{\sqrt{2}} \left[ (\bm{W}_\mu 
\cdot \bm{\hat x}) \pm i (\bm{W}_\mu \cdot \bm{\hat y}) \right]\,, \cr
Z^0_\mu &=& (\bm{W}_\mu \cdot \bm{\hat z}) \cos \theta_W 
- B_\mu \sin \theta_W \,,\cr
A_\mu &=&  B_\mu \cos \theta_W  
+ (\bm{W}_\mu \cdot \bm{\hat z})\sin \theta_W \,,
\end{eqnarray}
where $\bm{\hat x}$, $\bm{\hat y}$, and $\bm{\hat z}$ denote unit
vectors in isospin space and $\theta_W = \arctan(g'/g)$ is the weak
mixing angle.  The $W^\pm_\mu$ fields have mass $m_W=gv/\sqrt{2}$ and
electric charge $\pm e = \pm g'\cos \theta_W$, the $Z^0_\mu$ field has
mass $m_Z = m_W/\cos \theta_W$ and zero electric charge, and the
photon field $A_\mu$ has zero mass and zero electric charge.  The only
other physical degree of freedom in the theory is the magnitude of
the Higgs field, with mass $m_H=2v\sqrt{\lambda}$ and zero electric charge.

\section{Lattice Theory}

To analyze the classical equations of motion numerically,
we use the standard Wilsonian approach \cite{Wilson} for
lattice gauge fields (for a review see \cite{SmitBook}),
adapted to Minkowski space evolution as in 
\cite{Shaposhnikov,SmitSim,RajantieSim}.  The $U(1)$ and $SU(2)$
gauge fields live on the links of the lattice and the Higgs field
lives at the lattice sites.  We use a regular lattice with spacing
$\Delta x$ and determine the values of the fields at time $t_+ =
t+\Delta t$ based on their values at times $t$ and $t_- = t-\Delta t$.  
Throughout, we will use the same notation and conventions as
\cite{oscsm}.

We associate the Wilson line
\begin{equation}
U_j^p = e^{i g' B_j^p \Delta x /2}  
e^{i g \bm{W}_j^p\cdot\bm{\tau} \Delta x/2}
\label{eq:Wilson}
\end{equation}
with the link emanating from lattice site $p$ in the positive $j^{\rm
th}$ direction.  We define the Wilson line for the link emanating from
lattice site $p$ in the negative $j^{\rm th}$ direction to be the
adjoint of the corresponding Wilson line emanating in the positive
direction from the neighboring site, $U_{-j}^p = (U_j^{p-j})^\dagger$,
where the notation $p \pm j$ indicates the adjacent lattice site to
$p$, displaced from $p$ in direction $\pm j$.  At the edges of the
lattice we use periodic boundary conditions.

The equation of motion for the Higgs field at site $p$ is
\begin{equation}
\Phi^p(t_+) = 2 \Phi^p\currtime - \Phi^p(t_-) + \Delta t^2 \ddot
\Phi^p \currtime \,,
\end{equation}
where
\begin{equation}
\ddot \Phi^p \currtime = \sum_{j=\pm x,\pm y, \pm z}
\frac{U_j^p\currtime \Phi^{p+j}\currtime - \Phi^p\currtime}{\Delta
x^2} + 2\lambda\left(v^2 - |\Phi^p\currtime|^2\right)\Phi^p\currtime \,.
\end{equation}
For the gauge fields, we have
\begin{eqnarray}
U_j^p(t_+) =
\exp\left[\log U_j^p\currtime U_j^p(t_-)^\dagger
-\left( 
\sum_{j'\neq j} 
\frac{\log U^p_{\square(j,j')}\currtime + 
\log U^p_{\square(j,-j')}\currtime}{\Delta x^2}
+ \frac{i\Delta x}{2} (g' J_j^p + g \bm{J}_j^p \cdot \bm{\tau} )
\right)\Delta t^2 \right] U_j^p\currtime \,,
\end{eqnarray}
where $U^p_{\square(j,j')}\currtime = U_j^p\currtime
U_{j'}^{p+j}\currtime U_{-j}^{p+j+j'}\currtime
U_{-j'}^{p+j'}\currtime$ and
\begin{equation}
J_j^p = g' {\rm \, Im \,}
\frac{\Phi^p\currtime^\dagger U_{j}^p\currtime
\Phi^{p+j}\currtime}{\Delta x}\,, \qquad
\bm{J}_j^p = g {\rm \, Im \,} 
\frac{\Phi^p\currtime^\dagger  \bm{\tau} U_{j}^p\currtime
\Phi^{p+j}\currtime}{\Delta x} \,,
\end{equation}
are the gauge currents. Here we have defined the logarithm of 
a $2\times 2$ matrix in the form of Eq.\ (\ref{eq:Wilson}) as
\begin{equation}
\log U_j^p = \frac{i\Delta x}{2} (g' B_j^p + g \bm{W}_j^p\cdot\bm{\tau}) \,,
\end{equation}
which gives the more familiar gauge fields in terms of the link
variables.  We note that $\log XY \neq \log X + \log Y$ when the matrices
do not commute.

The $U(1)$ and $SU(2)$ matrices in Eq.\ ({\ref{eq:Wilson}}) are stored
separately in the numerical code.  To represent the $U(1)$ matrix
$U_1=e^{i\theta}$, just the real quantity $\theta=g' B_j^p \Delta x
/2$ is actually stored.  Any $SU(2)$ matrix can be written as
\begin{equation}
U_2=\pmatrix{x_1 & x_2 \cr -x_2^* & x_1^*}\,,
\end{equation}
so only the two complex elements of the top row need to be stored.
(This representation is redundant, since $|x_1|^2 + |x_2|^2 = 1$, but
more efficient computationally than storing three real quantities and
reconstructing the fourth.)  The logarithms and exponentials needed to
convert between the group and the algebra can be computed efficiently
using
\begin{equation}
U_2 = e^{i \theta \bm{\hat n} \cdot \bm{\vec \tau}} =
\cos \theta + i \bm{\hat n}\cdot \bm{\vec \tau} \sin \theta
= \pmatrix{
\cos \theta + i \bm{\hat n}_z \sin \theta &
i\bm{\hat n}_x \sin \theta + \bm{\hat n}_y \sin \theta \cr
i\bm{\hat n}_x \sin \theta - \bm{\hat n}_y \sin \theta &
\cos \theta - i \bm{\hat n}_z \sin \theta} \,,
\label{logformula}
\end{equation}
where $\bm{\hat n}$ is a unit vector and the link matrices have
$\bm{\hat n} \theta = \bm{W}_j^p g \Delta x /2$.

We note that this discretization differs slightly from the standard
approach used in \cite{Shaposhnikov,SmitSim,RajantieSim}.  In our
language, their discretization is equivalent to replacing $\sin\theta
\rightarrow \theta$ and $\cos\theta \rightarrow \sqrt{1-
\theta^2}$ when computing both the logarithm and the corresponding
exponential.  While the approach we are using corresponds a little
more directly to the continuum equations, any differences are of
higher order in the lattice spacing.  Numerical experiments show that
their approach yields completely equivalent results, and is somewhat
more efficient computationally, since it avoids the need to compute
trigonometric functions in this conversion.

The energy density at $p$ is then
\begin{eqnarray}
u^p\currtime &=& \frac{1}{2} \sum_{j=x, y, z} \left[ 
\frac{\left\|\exp \left(\log U_j^p(t_+) - \log U_j^p(t_-) \right)\right\|^2}
{(2 \Delta t)^2} + \sum_{j'> j} \frac{\left\|
U^p_{\square(j,j')}\currtime\right\|^2}{\Delta x^2}\right]
\cr &&
+ \frac{\left|\Phi^{p}(t_+) - \Phi^{p}(t_-) \right|^2} {(2 \Delta t)^2}
+\sum_{j=x, y, z} \frac{\left|U_j^p\currtime\Phi^{p+j}\currtime 
- \Phi^{p}\currtime \right|^2}{\Delta x^2} 
+ \lambda\left(|\Phi^p|^2 - v^2\right)^2 \,,
\end{eqnarray}
whose integral over the whole lattice is conserved.  Here we have
defined 
\begin{equation}
\left\|U_j^p\right\|^2 =
\frac{\left|{\rm Tr\,} \log U_j^p\right|^2}{g'^2 \Delta x^2} +
\frac{\left({\rm Tr\,} \bm{\tau} \log U_j^p \right)^\dagger \cdot \left(
{\rm Tr\,} \bm{\tau} \log U_j^p \right)}{g^2 \Delta x^2}
= |B_j^p|^2 + \bm{W}_j^p\cdot\bm{W}_j^p
\end{equation}
for any $U(2)$ link matrix.

At every lattice point, Gauss's Law,
\begin{eqnarray}
\sum_{j=x, y, z}
\frac{\log U_j^p(t_+)    U_j^p\currtime^\dagger
 +    \log U_{-j}^p(t_+) U_{-j}^p\currtime^\dagger}
{2 i \Delta x^2 \Delta t}
- \left(g' J_0^p + g \bm{J}_0^p \cdot \bm{\tau}\right) = 0\,,
\label{Gauss}
\end{eqnarray}
is also maintained throughout the evolution, where the charge
densities are given by
\begin{equation}
J_0 = g' {\rm \, Im \,} \left(\frac{\Phi^{p}(t_+) -
\Phi^{p}\currtime} {\Delta t} \right)^\dagger \Phi^{p}\currtime \,, \qquad
\bm{J}_0 = g {\rm \, Im \,} \left(\frac{\Phi^{p}(t_+) -
\Phi^{p}\currtime} {\Delta t} \right)^\dagger \bm{\tau}
\Phi^{p}\currtime \,.
\end{equation}
This requirement will provide a stringent check on the correctness of
the numerical simulation.  Here we have computed Gauss's Law at time
$t+\Delta t/2$, which is obeyed exactly by the discrete equations of
motion for any time step and lattice spacing.  In \cite{oscsm},
Gauss's Law at time $t$ was used; it is only obeyed to order $\Delta
t^2$, but as a result it also provides a rough estimate of whether the
time step is small enough.

\section{Spherical Ansatz}

With the $U(1)$ field included, the grand spin symmetry of the
spherical ansatz used in \cite{oscillon} is broken and field
configurations will not maintain this symmetry under time evolution.
The continuum theory does still preserve invariance under grand spin
rotations around the $z$-axis, but the Cartesian lattice provides
a small breaking of all rotational symmetries.  As a result, field
configurations that start within the spherical ansatz are not
constrained to lie in any reduced ansatz at later times.  (We will
also demonstrate the oscillon's stability under explicitly
nonspherical deformations below.)  Nonetheless, because we will use
the spherical ansatz as a starting point to obtain our initial
conditions, it will be helpful to analyze it in more detail.  We will
see that the electroweak oscillon retains much of the structure it
inherits from these initial conditions.

For our choice of gauge, the spherical ansatz takes the
form \cite{spherical}
\begin{eqnarray}
\bm{\tau} \cdot 
\bm{W}_j &=& 
\frac{1}{g}\left[
a_1(r,t) \bm{\tau}\cdot \bm{\hat r} \hat r_j +
\frac{\alpha(r,t)}{r}(\tau_j - {\bm{\tau}}\cdot\bm{\hat r}\hat r_j)
- \frac{\gamma(r,t)}{r}(\bm{\hat r} \times \bm{\tau})_j \right]\, , \cr
\Phi &=& \frac{1}{g} \left[\mu(r,t) - i \nu(r,t) \,
{\bm{\tau}}\cdot\bm{\hat r}\right]
\pmatrix{0 \cr 1} \,,
\label{ansatz}
\end{eqnarray}
where $\bm{r}$ is the position vector, $r=|\bm{r}|$ is the distance from
the origin, and $\bm{\hat r} = \bm{r}/r$ is the unit radial vector.
Configurations in this ansatz are then described by reduced fields
$a_1$, $\alpha$, $\gamma$, $\mu$, and $\nu$, all of which depend only on
$r$ and $t$.  The field definitions have been chosen so that the reduced
fields match those used in \cite{oscillon}, even though the conventions for
the three-dimensional theory used here are slightly different.  

These configurations are in the grand spin zero channel, meaning they are
symmetric under simultaneous rotations in space and isospin.  The
gauge field $\bm{W}_j$ has isospin $i=1$ and internal angular
momentum $s=1$.  These two spins can be coupled together to yield total
generalized angular momentum $0$, $1$, and $2$.  To obtain grand spin
$G=0$, these combinations must then be coupled with equal orbital
angular momenta $\ell=0$, $\ell=1$, and $\ell=2$ respectively,
corresponding to monopole, dipole and quadrupole spatial distributions.
These three possibilities are reflected in Eq.\ (\ref{ansatz}) through
the three terms $\alpha(r,t)$, $\gamma(r,t)$, and $a_1(r,t)$.

We have written the Higgs field as a matrix times a fixed isospinor.
This matrix transforms under both the gauged $SU(2)_L$ and global
$SU(2)_R$ isospin transformations.  (We are only considering global
rotations in both cases, however.)  Under both transformations it has
isospin $i=1/2$, giving total isospin $i=0$ or $i=1$.  Since the Higgs
is a Lorentz scalar, with zero internal angular momentum, 
to obtain $G=0$ these two possibilities must be coupled to $\ell=0$
and $\ell=1$ respectively, corresponding to monopole and dipole
spatial distributions.  These possibilities appear in 
Eq.\ (\ref{ansatz}) as the terms $\mu(r,t)$ and $\nu(r,t)$. 

Although the spherical ansatz does not contain the $U(1)$ field, to
leading order in $\theta_W$ we can find the electric charge density
created by a spherical ansatz configuration for our choice of gauge
\cite{Ron},
\begin{equation}
J_0 = \frac{2 e z}{r^3 g^2} \left(\gamma \dot \alpha
- \alpha \dot \gamma \right) \,.
\label{sphericalJ}
\end{equation}
The charge shows a dipole structure centered on the $z$ axis --- as we
would expected since the electromagnetic interactions break the
grand spin symmetry by selecting the $z$ direction in isospin.
We note that this electric charge density is time independent 
(and thus does not radiate) if the $\alpha$ and $\gamma$ fields vary
sinusoidally in time with the same frequency.

\section{Numerical Simulation}

The initial conditions for the simulation are obtained starting from
an approximate functional fit to the solutions that were found in 
$SU(2)$-Higgs theory using the spherical ansatz \cite{oscillon}.
These results, with slight modifications, provide the initial data for
the $\bm{W}_j$ and $\Phi$ fields, and the initial $B_j$ field is chosen to
vanish.  In order to guarantee that the initial configuration
obeys Gauss's Law in the full $SU(2)\times U(1)$ theory,
we generate the spherical ansatz fit at a point in the cycle where the
time derivatives are smallest, and then set all time derivatives to
zero.  We note that in pure $SU(2)$ Higgs-gauge theory, this
restriction would not be necessary, because even though an approximate
fit with nonvanishing time derivatives will not obey Gauss's Law, we
can restore Gauss's Law by adjusting $\Phi(t_+)$ slightly via an
$SU(2)$ transformation at each point,
\begin{equation}
\Phi_{\rm new}(t_+) = \left|\frac{\Phi_{\rm old}(t_+)}{\Phi\currtime}\right|
{\cal U}^p \Phi\currtime \,,
\end{equation}
with
\begin{equation}
{\cal U}^p = \exp\left[
\sum_{j=x, y, z}
\frac{\log U_j^p(t_+) U_j^p\currtime^\dagger
 + \log U_{-j}^p(t_+) U_{-j}^p\currtime^\dagger }
{g^2 \Delta x^2   |\Phi_{\rm old} (t_+)| |\Phi\currtime|/2}
\right]^\dagger \,.
\end{equation}
This procedure has been used successfully to reproduce
spherical ansatz solutions with nonvanishing time
derivatives at $t=0$ in a fully three-dimensional simulation of
pure $SU(2)$ Higgs-gauge theory, but it cannot be extended to the
$SU(2) \times U(1)$ theory because $\Phi$ carries both charges,
and thus cannot be adjusted to satisfy both constraints at once.
Therefore we will consider only initial conditions in which all fields
have zero time derivatives, so that Gauss's Law is trivially satisfied.

To construct the initial conditions, we begin from the spherical
ansatz form of Eq.\ (\ref{ansatz}).  We work in
units where $v=1/\sqrt{2}$. Since we are dealing with purely classical
dynamics, we can rescale the fields to fix the $SU(2)$ coupling constant at
$g = \sqrt{2}$, so that the $W^\pm$ mass is then
$m_W=gv/\sqrt{2}=1/\sqrt{2}$.  With this rescaling, we must also introduce
an overall factor of $g^2/g_W^2$ multiplying the total energy, where
$g_W=0.634$ is the true weak coupling constant.  (This factor
was incorrectly omitted in the original version of \cite{oscsm}.)
We choose $\lambda=1$, so that the Higgs mass is twice the $W^\pm$ mass,
$m_H=2v\sqrt{\lambda}=\sqrt{2}$.  Finally, we fix $g'=0.773$, so that
the ratio $g'/g$ matches its observed value and the $Z^0$ boson has
the correct mass. With these choices, one unit of energy is $114\ {\rm
GeV}$, one unit of time is $5.79 \times 10^{-27}\ {\rm sec}$, and one
unit of length is $1.74 \times 10^{-18}\ {\rm m}$.  In these units, we
take the following initial configuration for the radial fields,
\begin{eqnarray}
a_1(r) &=& \chi (0.117 \chi + 0.016 \chi r)
\left(\sech 2 \chi r\right)^{1/8} \,, \cr
\mu(r) &=& 1 - 0.138 \chi \sech \frac{\chi r}{6.75}\,, \cr
\nu(r) &=& 0.017 \chi r \sech \frac{\chi r}{5}\,, \cr
\alpha(r) &=& 0.117 \chi^2 r \sech\frac{\chi r}{8}\,, \cr
\gamma(r) &=& 0 \,,
\label{initial}
\end{eqnarray}
where the adjustable parameter $\chi$ allows us to include a
combined rescaling of the fields' amplitudes and $r$-dependence, as is
commonly used in a small amplitude analysis \cite{DHN,Hsu,smallamp}.
While $\chi=1$ gives an approximation to the spherical ansatz
solution of \cite{oscillon}, a slightly larger value appears to be
necessary for the configuration to settle into a stable solution in
the full $SU(2)\times U(1)$ model.  Here we will use $\chi=1.15$.
The first term in parentheses in the definition of $a_1(r)$ is scaled
with an additional $\chi$ so that it matches the coefficient of
$\alpha$, ensuring that $\alpha$, $a_1-\alpha/r$, $\gamma/r$, and
$\nu$ all vanish as $r \to 0$, as required for regularity of the
fields at the origin.  Within the spherical ansatz simulation, these
initial conditions converge to a long-lived oscillon in the pure
$SU(2)$-Higgs theory, which is never observed to decay.  
As a check of the numerical calculation, the full three-dimensional
simulation agrees with the spherical ansatz simulation when the $U(1)$
interaction is turned off.

Although initial conditions of this form do settle into stable
oscillon configurations in the $SU(2)\times U(1)$ theory, it is
helpful to make a minor modification to them that is outside the
spherical ansatz:  setting the $\tau_z$-component of $\bm{W}_j$ to
zero brings the initial conditions significantly closer to the
localized solution that the fields ultimately converge to.  While we
obtain an equivalent oscillon solution in both cases, this
modification reduces the energy shed as the oscillon forms.  Doing so
provides a significant technical benefit, because the radiation
emitted as the configuration settles into the oscillon solution can
wrap around the periodic boundary conditions, return to the region of
the oscillon, and potentially destabilize it.  To avoid this problem,
the energy density in this radiation, which spreads throughout the
volume of the simulation, must be small compared to the oscillon's
energy density.  As long as the lattice volume is large enough
compared to the oscillon size,  this radiation is sufficiently diffuse
that it does not affect the oscillon's evolution.  We use a lattice of
size $L=144$ on a side in natural units, which is more than enough to
satisfy this criterion.  For $L \gtrsim 100$, changing the lattice
size simply changes the pattern of noise caused by electromagnetic
radiation superimposed on the oscillon region, but does not affect
oscillon properties or stability.  We can therefore be certain that
there is no coherent structure to this unphysical radiation that could
possibly be necessary for the oscillon's stability.   Its only
potential effect is to destabilize the oscillon, and it only does so
when artificially concentrated by a small lattice (e.g. of
size $L < 100$).  In numerical experiments, these destabilization
effects are actually much weaker in the electroweak model than in pure
scalar or $SU(2)$ Higgs-gauge models, because in the electroweak model
the radiated energy ends up almost entirely in the electromagnetic
field, while the oscillon arranges itself to be electrically neutral.
For this reason, it is not necessary to use absorptive techniques such
as adiabatic damping \cite{2d1} or an expanding background
\cite{oscex}, although both have been applied successfully to this
problem as well.  However, clearly it is helpful to adjust the initial
conditions to be as close as possible to the true oscillon
configuration, to minimize the amount of unwanted energy
emitted as the configuration settles into the oscillon solution, and
therefore limit the numerical costs associated with a larger lattice.

\begin{figure}[htbp]
\includegraphics[width=0.5\linewidth]{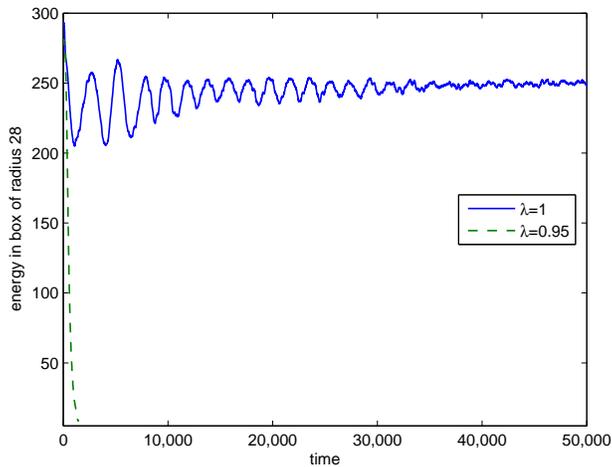}
\caption{Energy in a spherical box of radius 28 as a function of
time in natural units.  The initial conditions are given by the
modified spherical ansatz form given in the text, in which 
the $\tau_z$ component of the gauge field is set to zero, with 
$\chi=1.15$.  Two values of the Higgs self-coupling $\lambda$ are
shown.  For $\lambda=1$, the masses of the Higgs and $W$ fields are in
the $2:1$ ratio needed for oscillon formation and the solution remains
localized throughout the simulation.   Here one unit of energy is
$114\ {\rm GeV}$, one unit of time is $5.79 \times 10^{-27}\ {\rm
sec}$, and one unit of length is $1.74 \times 10^{-18}\ {\rm m}$,
giving a total energy of roughly 30 TeV within the box radius of
roughly 0.05 fm.  A transient beat pattern is also visible.  For
$\lambda=0.95$, the mass ratio is $1.95:1$.  In that case, there is no
stable object and the energy quickly disperses.}
\label{fig:main}
\end{figure}

Starting from the modified spherical ansatz initial conditions, we let
the system evolve for as long as is practical numerically, and see no
sign of oscillon decay.  We use lattice spacing $\Delta x = 0.75$,
though $\Delta x = 0.625$ and $\Delta x = 0.25$ were verified to give
completely equivalent results in correspondingly smaller tests.  The
time step is $\Delta t=0.1$.  Time steps of $0.05$ and $0.025$ also
gave equivalent results, although in this case one must take into
account the fact that this change also slightly alters the initial
conditions:  To set the initial time derivatives to zero, the
simulation sets the first two time slices equal.  Changing the time
step thus changes the time at which the field configuration matches
its value at $t=0$, representing a slight perturbation of the initial
conditions.  This change slightly alters the initial transient
behavior as the fields approach the oscillon, but these differences
quickly disappear and the simulations approach equivalent oscillon
configurations.

Total energy is conserved to a few parts in $10^3$
for $\Delta t= 0.1$, which improves with $\Delta t^2$ as expected for
our second-order algorithm.  We check Gauss's Law by monitoring the
left-hand side of Eq.\ (\ref{Gauss}), which we verify vanishes to
machine precision throughout the simulation.\footnote{One can instead
evaluate Gauss's Law at time $t$ instead of $t+\Delta t/2$ as in
\cite{oscsm}.  In that case, we square the left-hand side of Eq.\
(\ref{Gauss}), take its trace, and then take the square root of the
result.  For a typical run with $\Delta t = 0.1$, the integral of this
quantity over the lattice never exceeds $0.025$ and shows no upward
trend over time.  For smaller $\Delta t$, we see the expected ${\cal
O}(\Delta t^2)$ improvement in this result.}  It is necessary,
however, to use double precision to avoid gradual degradation in this
result.  For the parameters as given above, a run to time $10,000$
takes roughly $40$ hours using $24$ parallel processes, each running
on a $2\ {\rm GHz}$ Opteron processor core.\footnote{The parallel C++
code used for these simulations is available from 
{\tt http://community.middlebury.edu/\~{}ngraham}.} 

\begin{figure}
\includegraphics[width=0.45\linewidth]{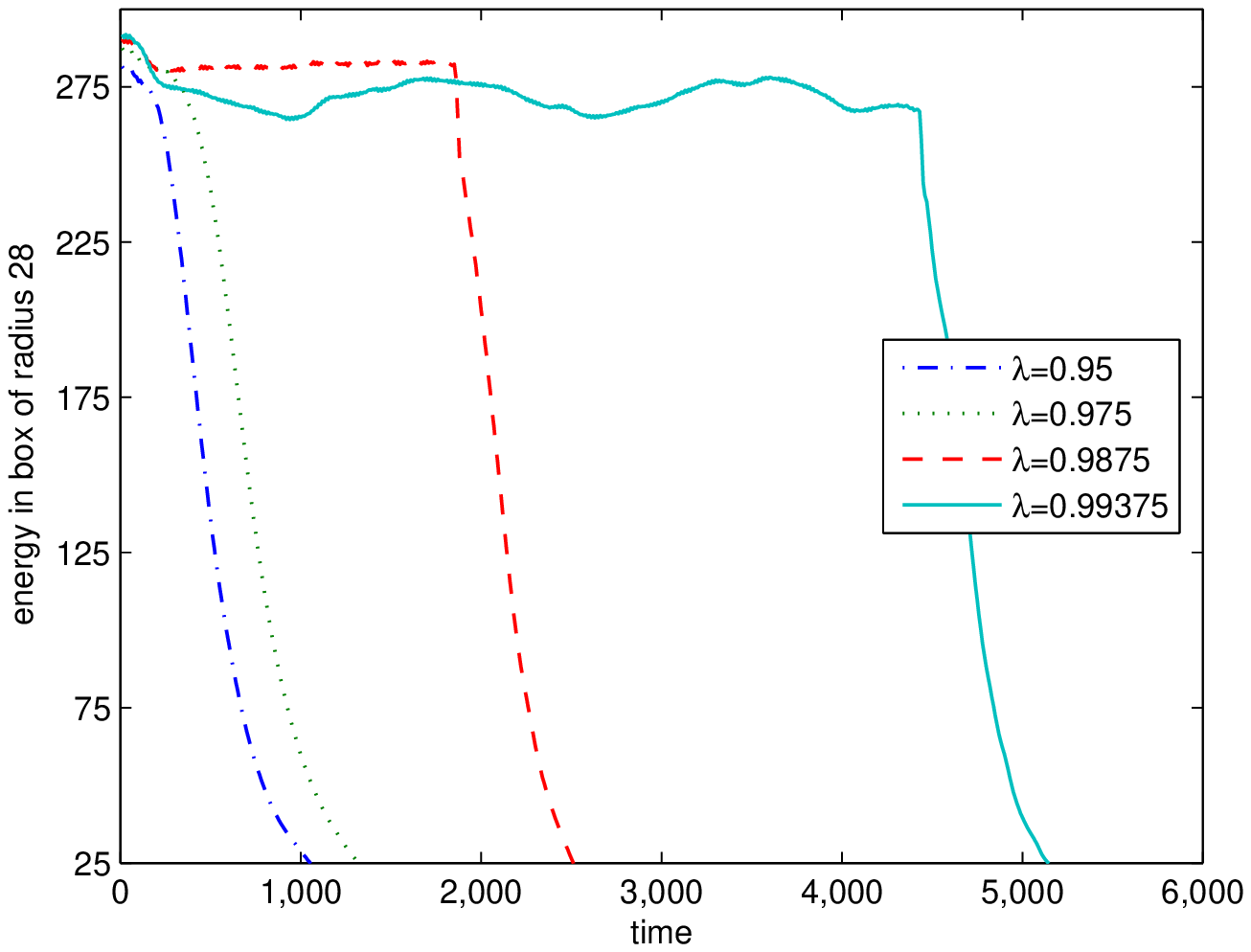}
\includegraphics[width=0.45\linewidth]{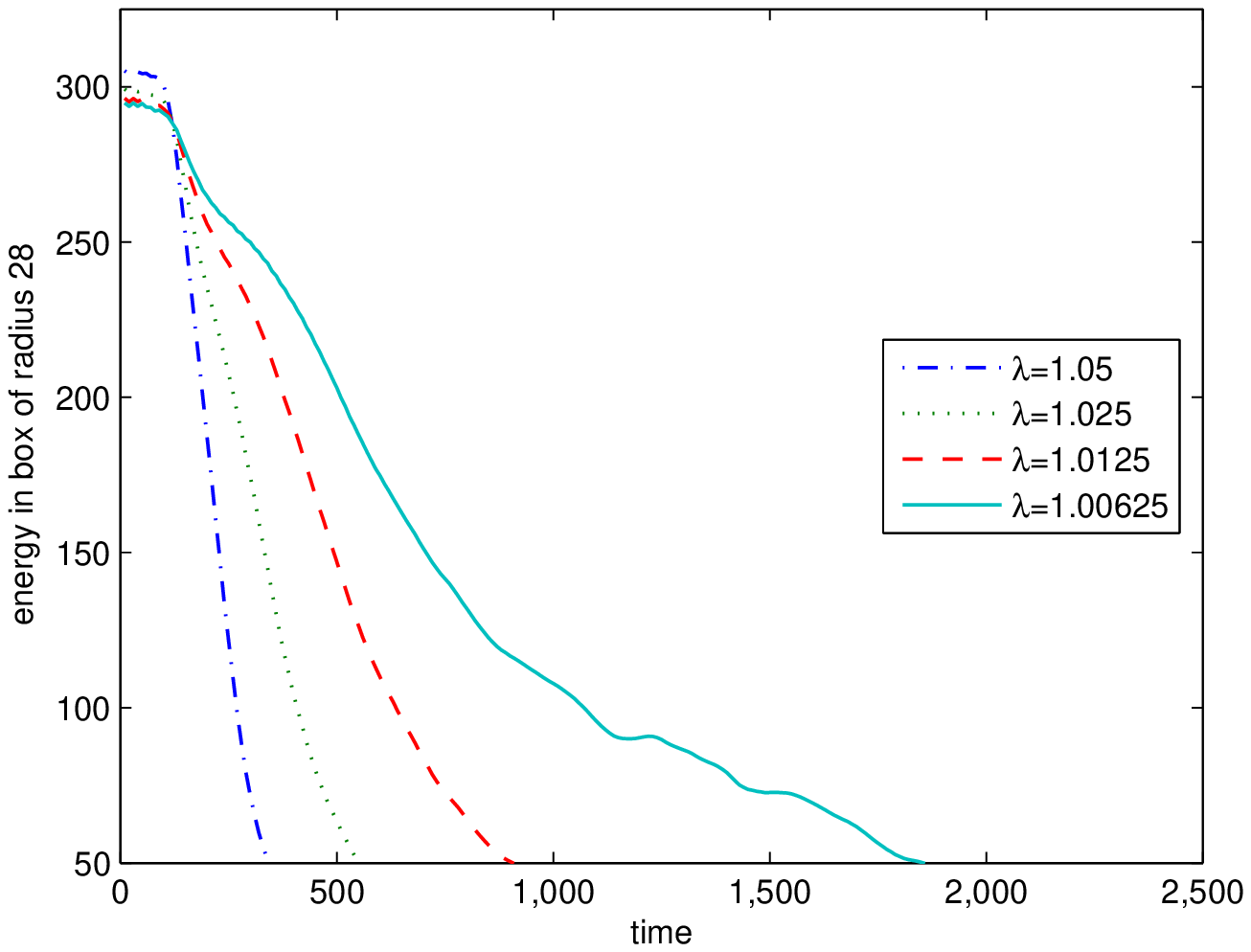}
\caption{Energy in the spherical box for a variety of values of
$\lambda$.  For $\lambda=1$, the Higgs mass is twice the $W^\pm$
mass and no decay is observed.  When the Higgs mass is just below this
value, we see a region of meta-stability.  For $\lambda < 1$, the
fields decay by first collapsing inward before dispersing, while for
$\lambda > 1$ the fields simply disperse outward.
}
\label{fig:massratio}
\end{figure}

\begin{figure}
\includegraphics[width=0.45\linewidth]{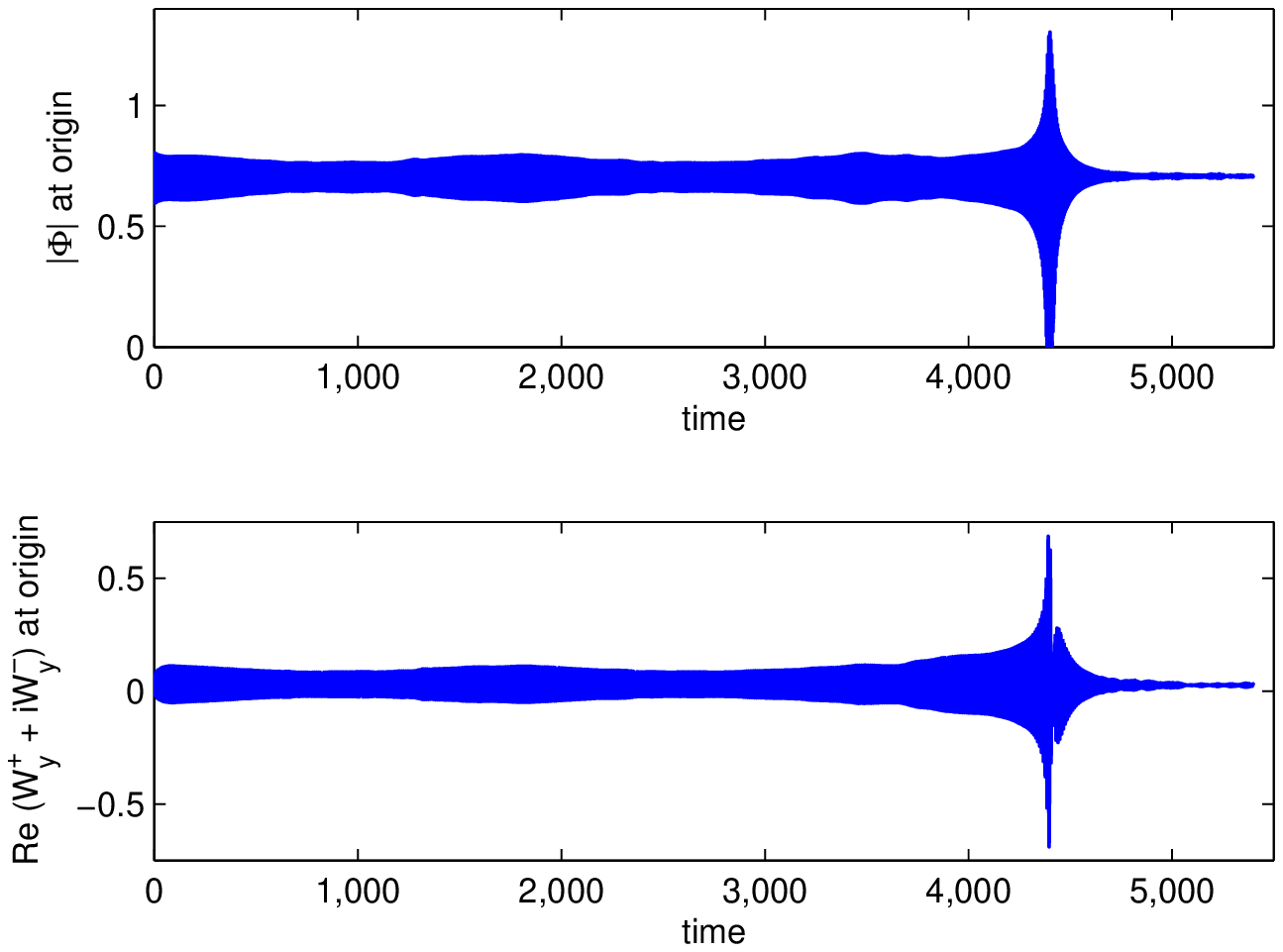}
\includegraphics[width=0.45\linewidth]{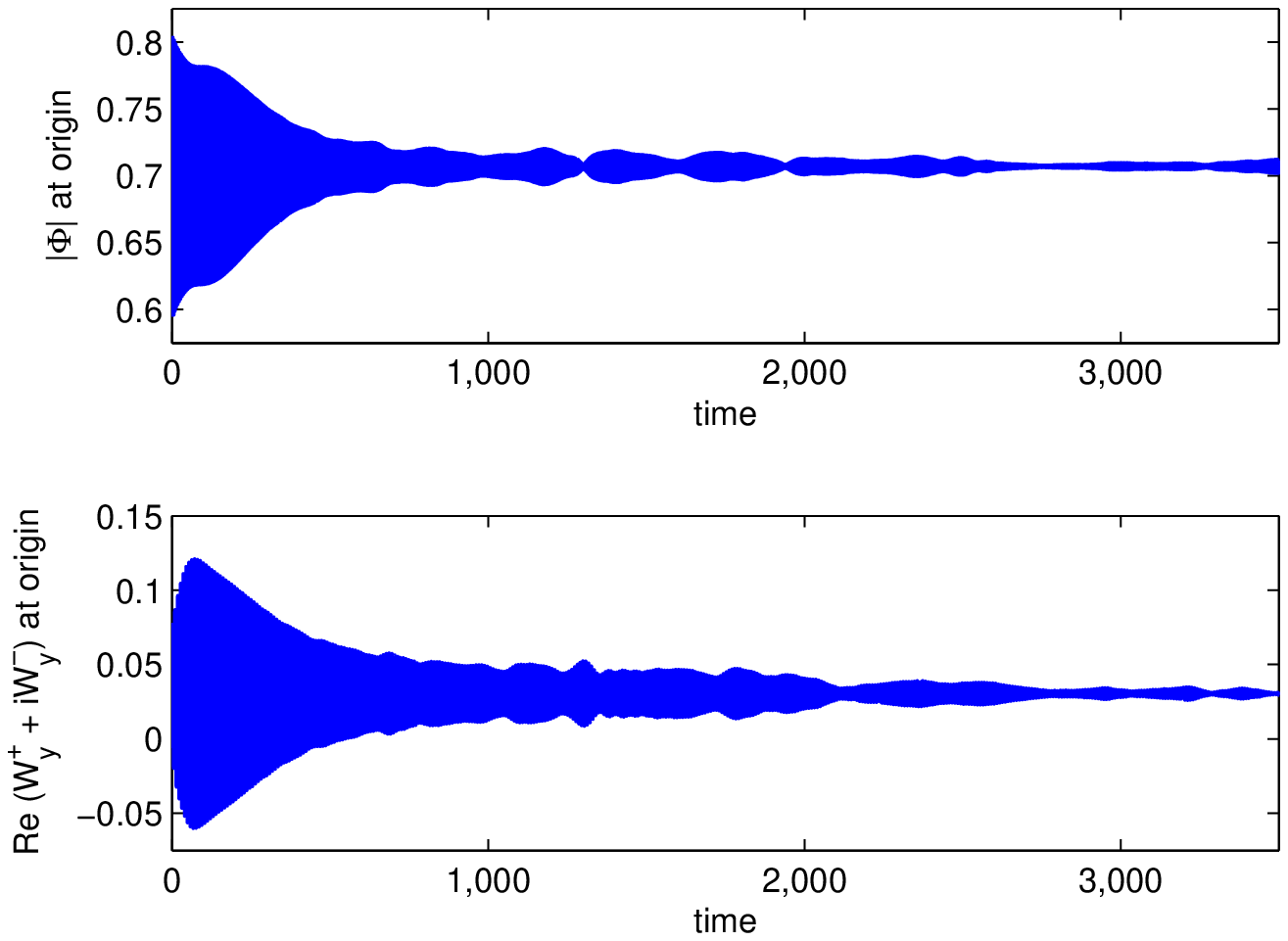}
\caption{
Decay of the oscillon for $\lambda \neq 1$.  One of the gauge fields
and the magnitude of the Higgs field at the origin are shown as
functions of time.  In the left panel $\lambda=0.99375$ and the
oscillon decays by collapsing inwards, creating a large amplitude
fluctuation at the origin  before dispersing.  In the right, panel,
$\lambda=1.00625$, and the oscillon decays by expanding outwards.
}
\label{fig:decay}
\end{figure}

\begin{figure}[htbp]
\includegraphics[width=0.5\linewidth]{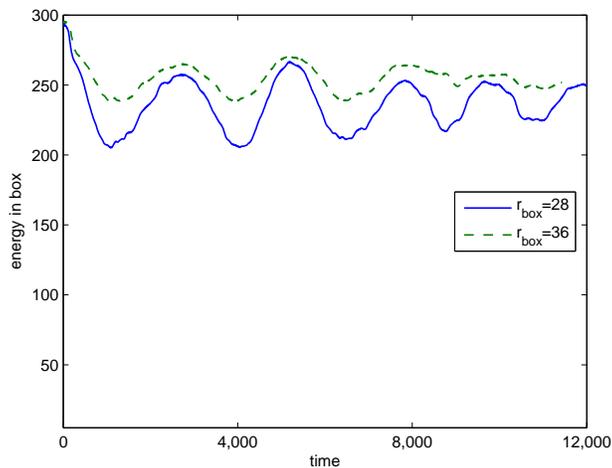}
\caption{Energy in the spherical box for two different box radii in
the simulation of Fig.\ \ref{fig:main}, with $\lambda=1$.
The transient beat pattern represents a ``breathing'' perturbation in
which the oscillon stretches and compresses slightly.  For the larger
box size, less energy flows in and out of the box during this process,
and so the observed beat amplitude is smaller.
}
\label{fig:box}
\end{figure}

Fig.\ \ref{fig:main} shows the energy in a spherical box of
radius $28$ as the fields are evolved from these initial conditions.
When the Higgs mass is twice the $W^\pm$ mass, a small amount of
energy is initially emitted from the central region, with the rest
remaining localized for the length of the simulation.  If the masses
are not in this ratio, however, the initial configuration quickly
disperses.  Fig.\ \ref{fig:massratio} shows the growth in oscillon
lifetime as $\lambda$ approaches this critical value.  We see a region
of meta-stability when the Higgs mass is just below the $2:1$ ratio.  
For $\lambda < 1$, the fields first collapse toward the origin before
dispersing, while for $\lambda >1$ they simply spread outward.  This
behavior is shown in Fig.\ \ref{fig:decay}.  Other ``special'' ratios,
such as $m_H = 2 m_Z$, did not form stable objects from these initial
conditions.

The spherical box contains approximately 3\% of the total volume
available to the simulation.  Its radius has been chosen to be just
large enough to enclose nearly all of energy density associated
with the stable oscillon.  As a result of this choice, the $\lambda=1$
graph also shows a transient beat pattern.  It represents a
``breathing'' or ``ringing'' motion, in which the oscillon gradually
expands and contracts slightly over many periods, accompanied by a
corresponding modulation of the field amplitudes.  This process causes
a small amount of the oscillon's energy to move in and out of the box.
As we would expect, when a larger box size is used, the ``breathing''
is more completely contained within the box and the graph of the
energy in the box flattens out, as shown in Fig.\ \ref{fig:box}.
Similar beats appear in the $SU(2)$ spherical ansatz oscillon
\cite{oscillon}, but in the electroweak oscillon their amplitude
decays much more rapidly.

\begin{figure}[htbp]
\includegraphics[width=0.99\linewidth]{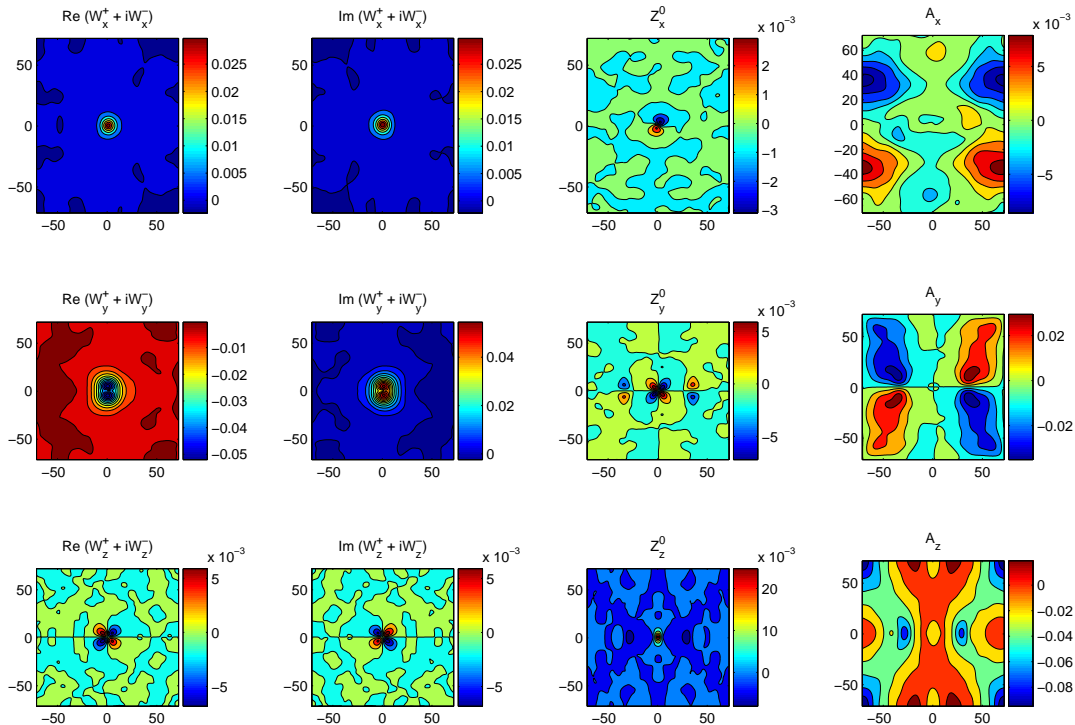}
\caption{A snapshot of the gauge fields in the $x=0$ plane 
for the simulation of Fig.\ \ref{fig:main} at time $t=50,000$.
Subscripts refer to spatial components.}
\label{fig:pot}
\end{figure}

\begin{figure}[htbp]
\includegraphics[width=0.99\linewidth]{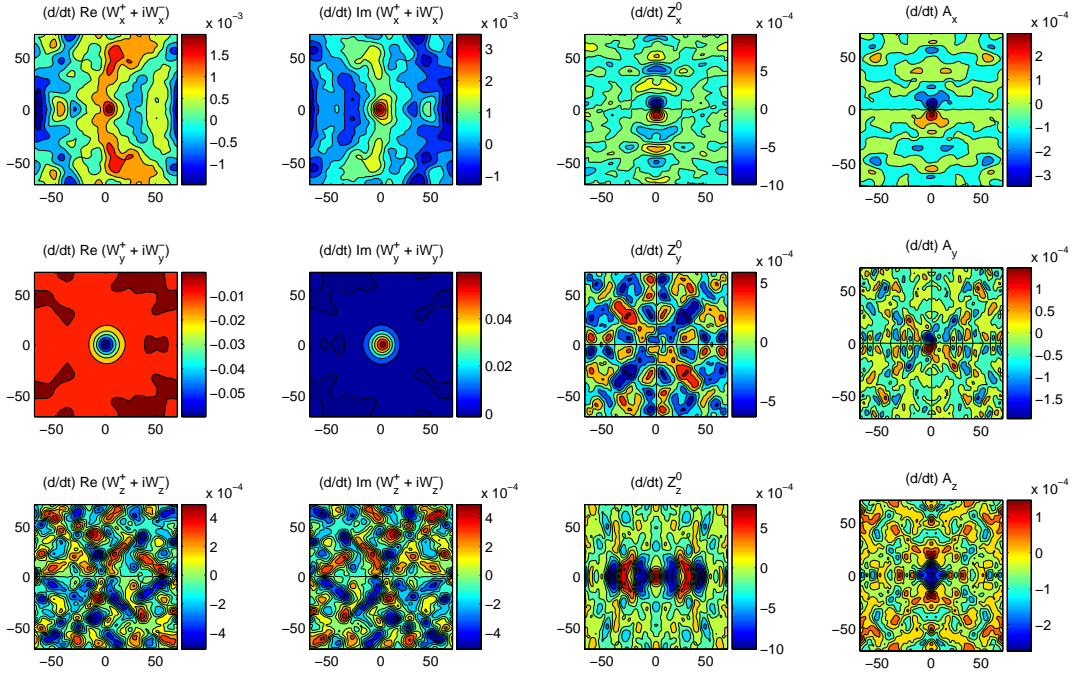}
\caption{A snapshot of the electric fields (time derivatives of the
gauge potentials) in the $x=0$ plane
for the simulation of Fig.\ \ref{fig:main} at time $t=50,000$.
Subscripts refer to spatial components.}
\label{fig:fields}
\end{figure}

\begin{figure}[htbp]
\includegraphics[width=0.99\linewidth]{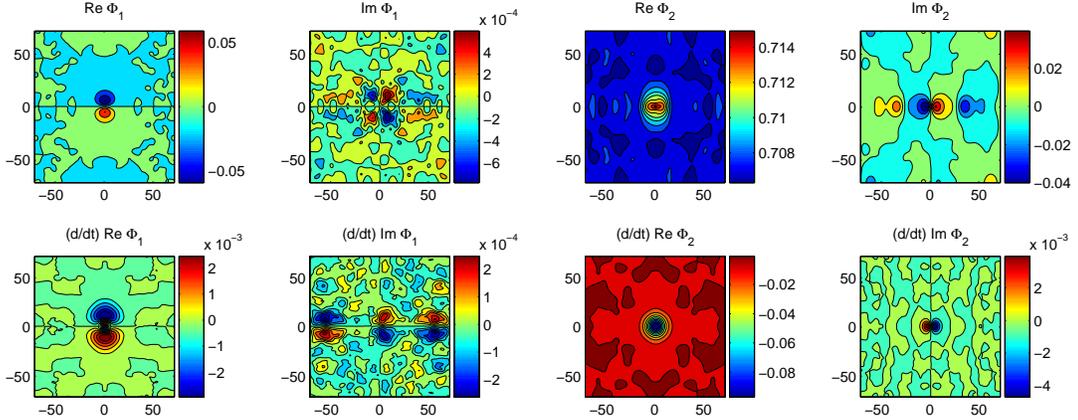}
\caption{A snapshot of the Higgs field and its time derivatives in the
$x=0$ plane for the simulation of Fig.\ \ref{fig:main} at time
$t=50,000$.  Subscripts refer to components of the Higgs field.}
\label{fig:phis}
\end{figure}

\begin{figure}[htbp]
\includegraphics[width=0.25\linewidth]{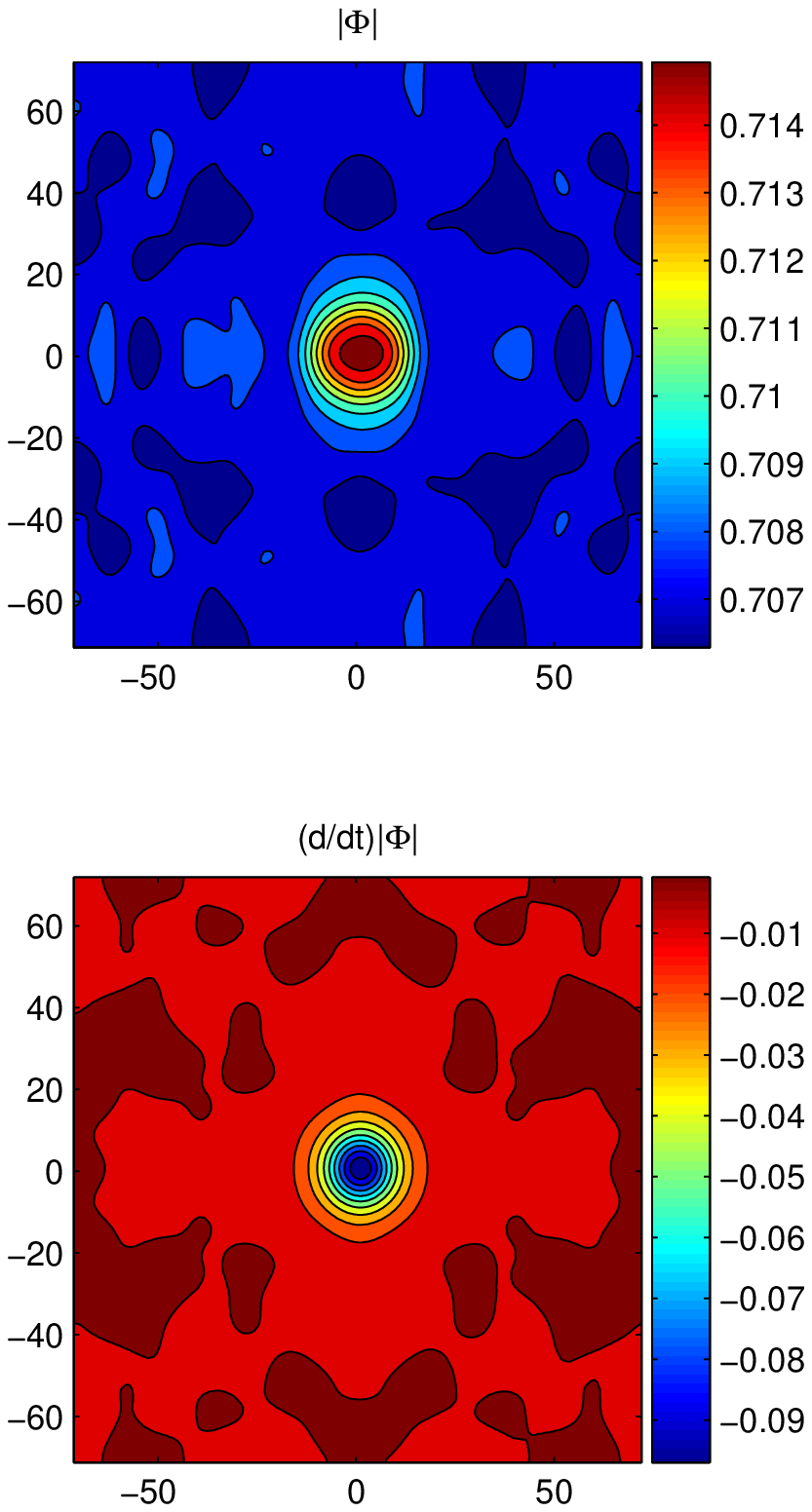}
\includegraphics[width=0.6\linewidth]{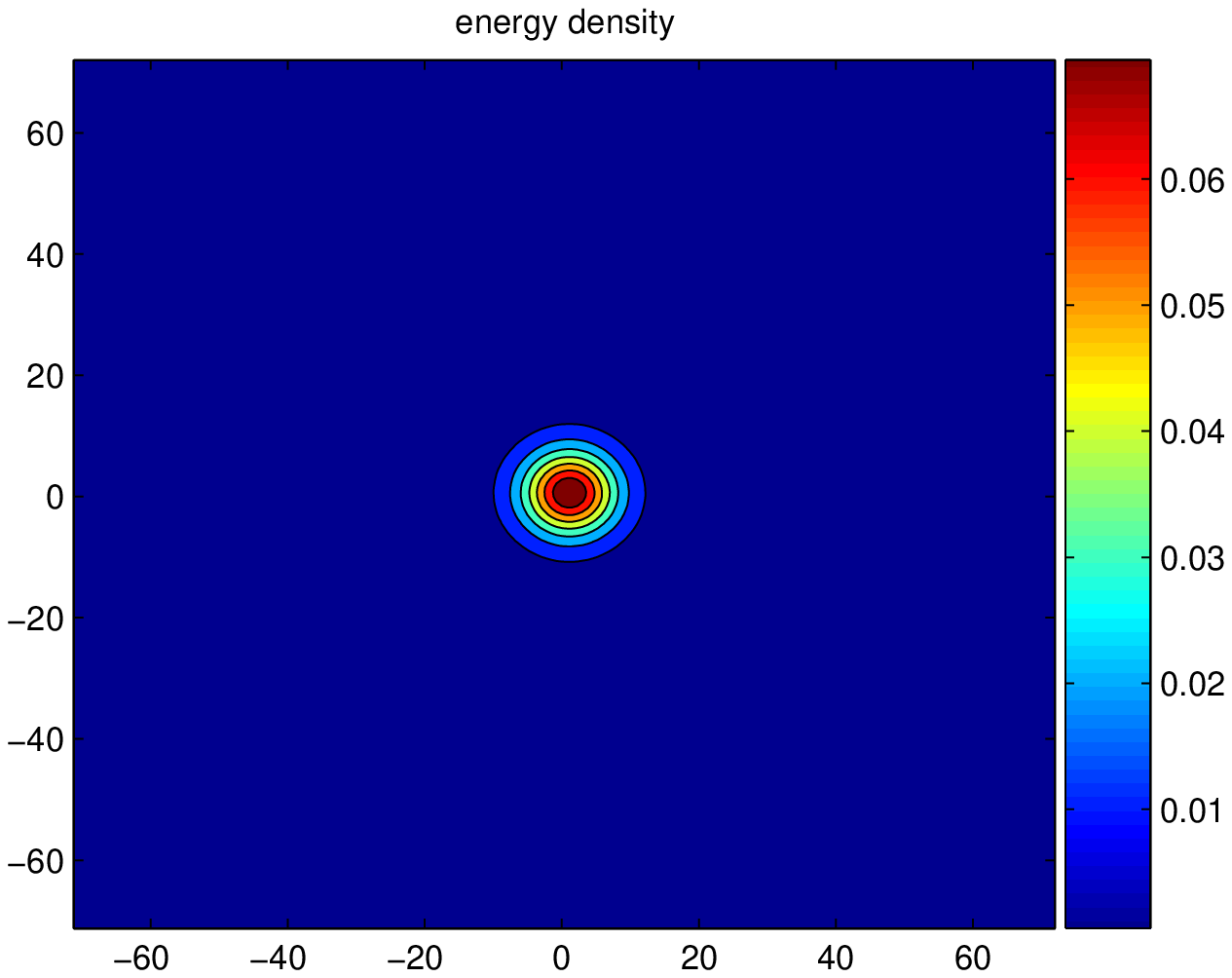}
\caption{
Left panel:  A snapshot of the magnitude of $\phi$ 
and its first time derivative in the $x=0$ plane
for the simulation of Fig.\ \ref{fig:main} at time $t=50,000$.
Right panel:  A snapshot of the energy density the $x=0$ plane
for the simulation of Fig.\ \ref{fig:main} at time $t=50,000$.
\label{fig:denmag}
}
\end{figure}

To illustrate the field configurations that make up the oscillon, we
graph the fields at time $t=50,000$ for the two-dimensional slice $x=0$.
Fig.\ \ref{fig:pot} shows the gauge field components.  It is most
illustrative to consider a linear superposition of the $W^\pm_j$
fields, as shown in the figure.  Fig.\ \ref{fig:fields} shows the
electric fields, which are given by the time derivatives of the gauge
fields for our choice of gauge. Fig.\ \ref{fig:phis} shows the
components of the Higgs field and  its first time derivative, and
Fig.\ \ref{fig:denmag} shows the magnitude of the Higgs field and the
first time derivative of this quantity, together with the total energy
density.  The oscillon is constructed primarily out of the lower
component of the Higgs field, the imaginary part of the upper
component of the Higgs field, the $x$ and $y$ spatial components of
the $W^\pm_j$ fields, and the $z$ spatial component of the $Z^0_j$
field.  We see the multipole structures we anticipated from the
spherical ansatz analysis.  The Higgs field contains monopole and
dipole fluctuations.  The photon field $A_j$ contains delocalized
background radiation that was emitted as the oscillon formed from the
initial conditions.  As we would expect from Eq.\ (\ref{sphericalJ}),
it has a dipole structure.  In the spherical ansatz, the $W^\pm_j$ and
$Z^0_j$ fields can potentially contain monopole, dipole, and
quadrupole components.  Here we see significant monopole and
quadrupole structures, but only a very small dipole component, which
appears in $Z^0_j$.  As a result, the electric charge we estimate from
Eq.\ (\ref{sphericalJ}) is very small, as is the true value from the
numerical simulation; the oscillon is decoupled from the electromagnetic
background.\footnote{While the multipole analysis is instructive as a
description of the field configuration, is is important to note that
because the oscillon has large spatial extent compared to its period of
oscillation, it is in exactly the domain where the standard multipole
expansion for the electromagnetic radiation emitted is invalid.}

\begin{figure}[htbp]
\includegraphics[width=0.45\linewidth]{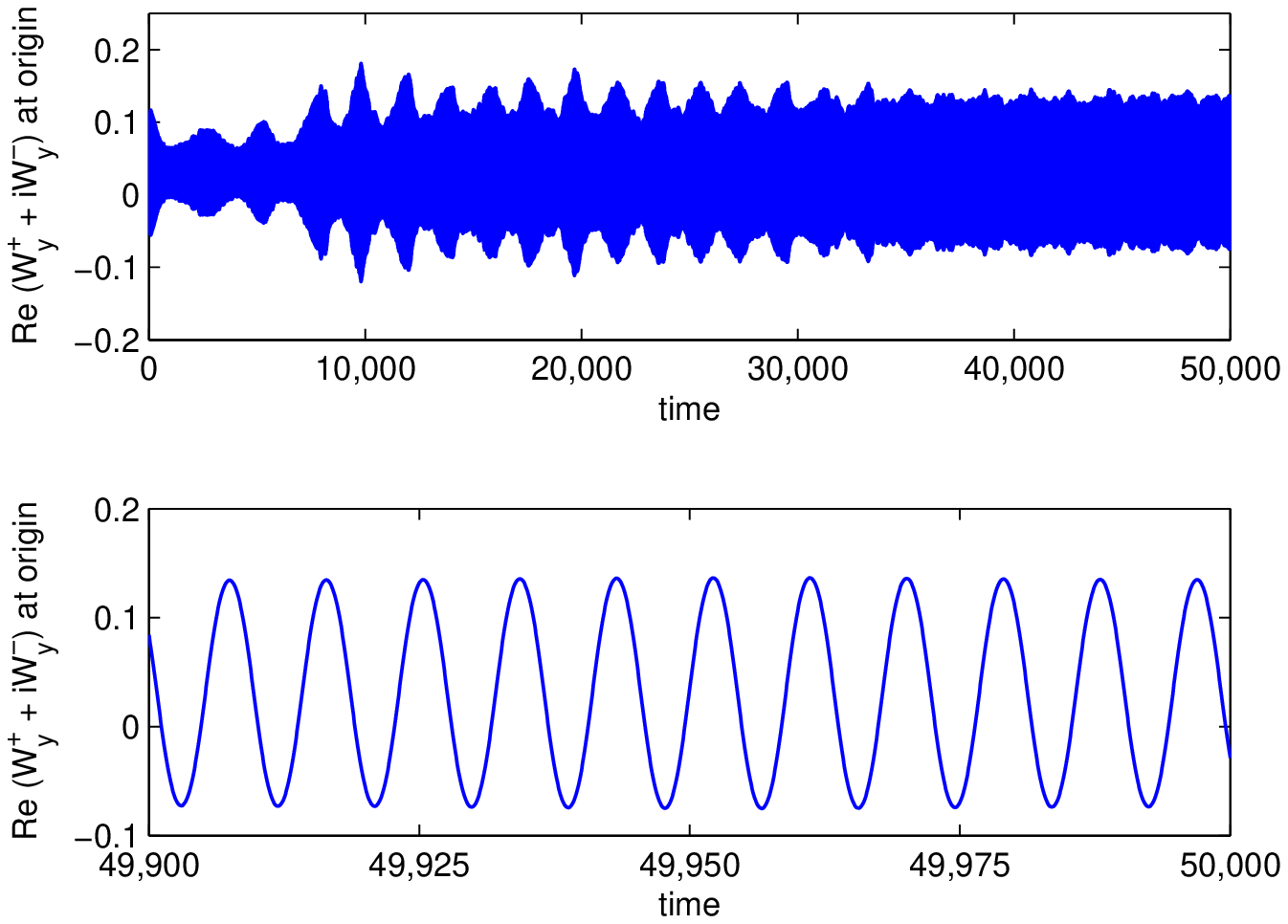}
\includegraphics[width=0.45\linewidth]{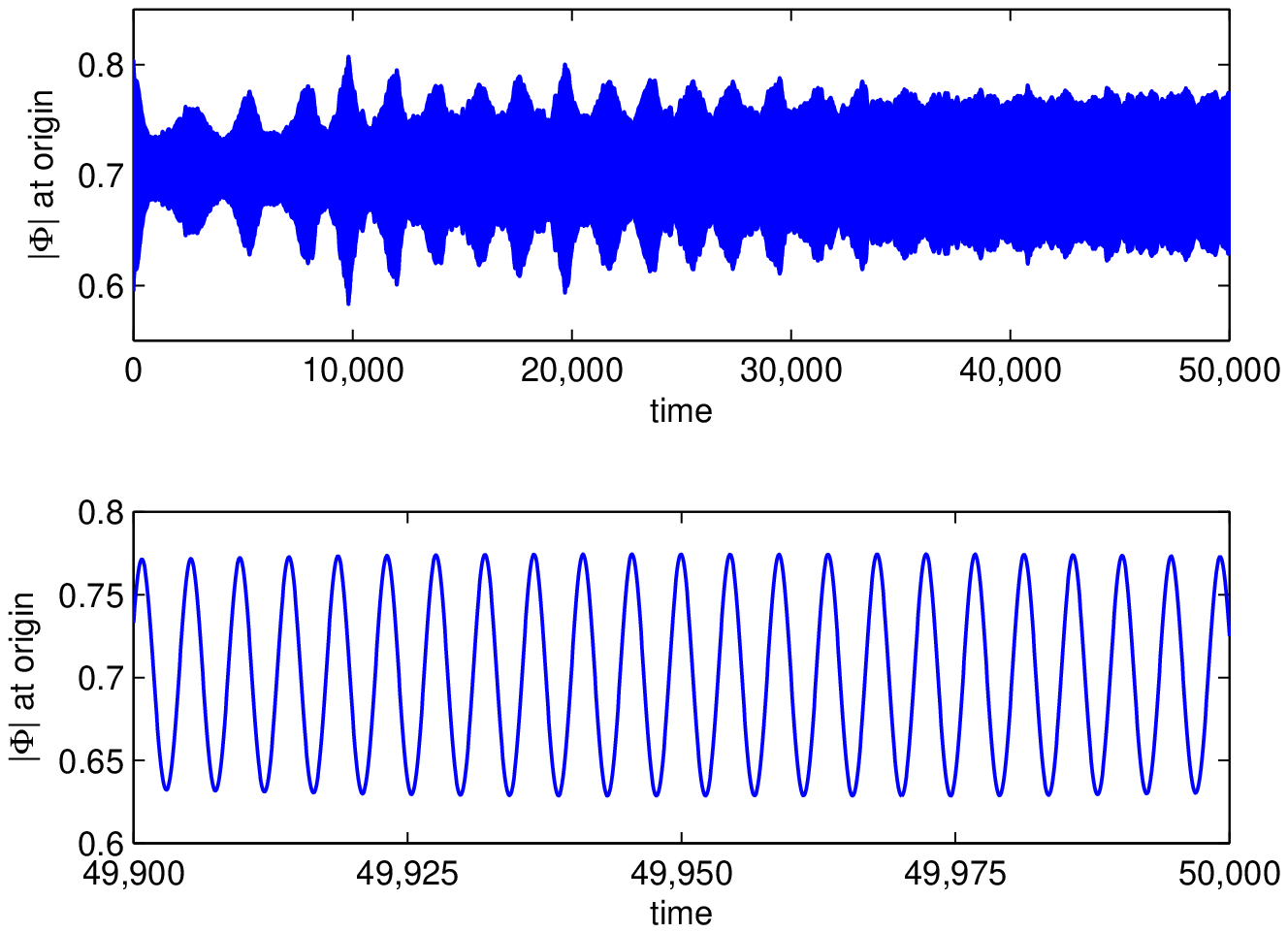}
\caption{
Oscillon fields at the origin as functions of time.  The left side
shows one component of the $SU(2)$ gauge field.  The upper graph shows
the full extent of the simulation.  On this scale, the individual
oscillations are too small to be seen.  Instead, we see the decaying
beat pattern from the transient ``breathing'' motion.  The lower graph
shows the oscillation of the field for a short time at the end of the
simulation (when the transient effects have decayed away).  The right
side shows the magnitude of $\Phi$ in the same way.  It oscillates with
fundamental frequency twice that of the gauge field.
}
\label{fig:origin}
\end{figure}

Each excited field oscillates at a frequency just below its mass.  In
our units, these oscillations have typical amplitude of order $0.1$
and typical radius of order $10$.  By comparing the total number of
cycles to the total time, we find $\omega_H = 1.404$ for the Higgs field
components and $\omega_W=0.702$ for the gauge field components.  These
properties are all very similar to the spherical ansatz oscillon.  They
are also consistent with a small-amplitude analysis, as described in
the Introduction, with $\epsilon$ of order $0.1$.  In Fig.\
\ref{fig:origin}, oscillon fields at the origin are shown as
functions of time.  The fundamental oscillation of each field is
modulated by the decaying beat pattern.

\begin{figure}[htbp]
\includegraphics[width=0.5\linewidth]{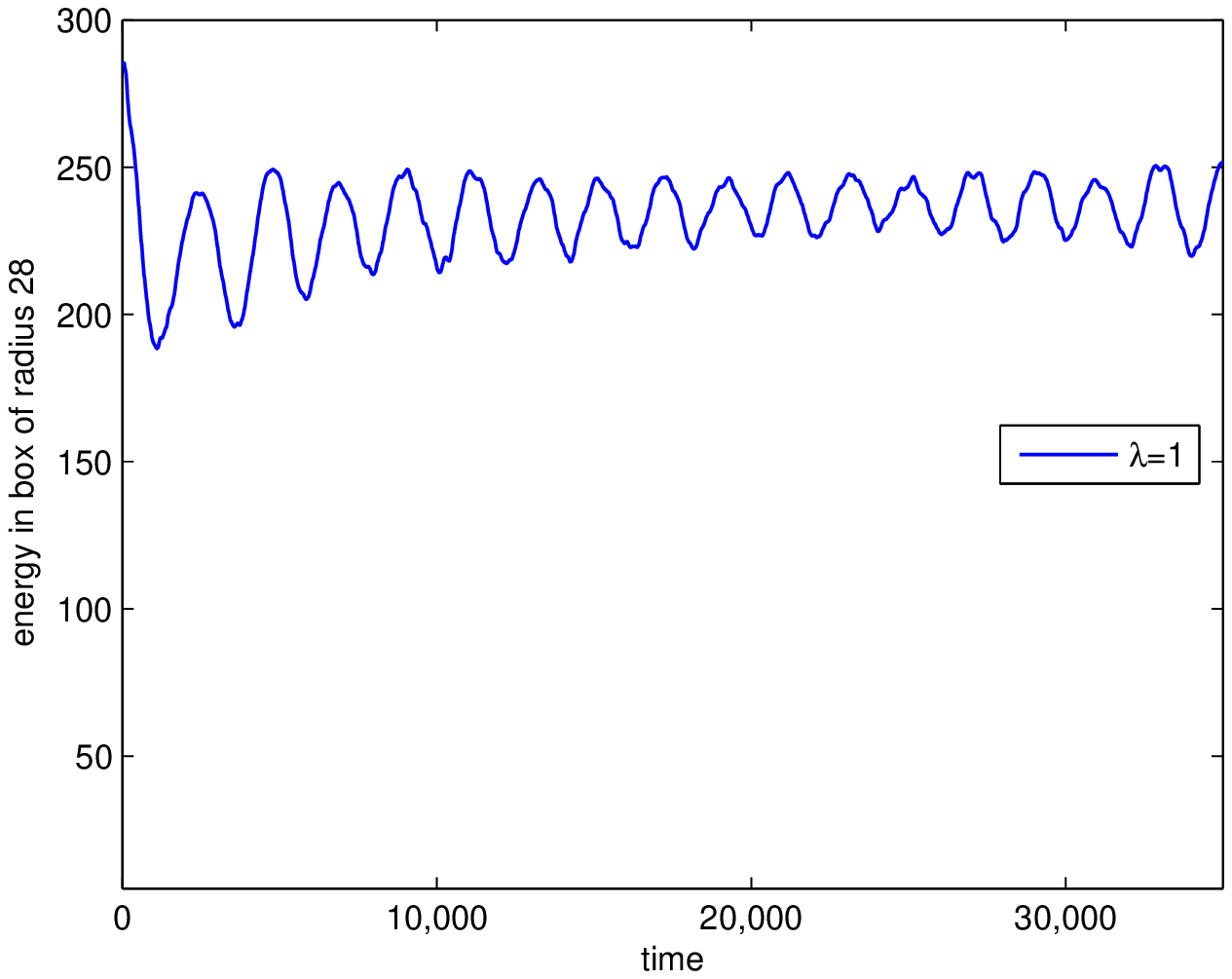}
\caption{
Energy in a box of radius $28$ as in Fig.\ \ref{fig:main}, but with
initial conditions that have been deformed to break rotational
symmetry.  The spatial coordinate
$\bm{r} = x \bm{\hat x} + y \bm{\hat y} + z \bm{\hat z}$
has been everywhere replaced by  
$\bm{r'} = 0.98 x \bm{\hat x} + 1.02 y \bm{\hat y} + 0.97 z \bm{\hat z}$ 
and similarly $r$ and $\bm{\hat r}$ have been replaced by 
$r' = |\bm{r'}|$ and $r' = \bm{x'}/r'$.  As a further check of the
numerics, this run also uses a smaller time step, $\Delta t = 0.05$.
Except for these modifications, the simulation is the same as in
Fig.\ \ref{fig:main}.}
\label{fig:def}
\end{figure}

The oscillon we have seen is not significantly altered by
small perturbations of the initial conditions.  As an example, in
Fig.\ \ref{fig:def} we show the results of a run in which the
rotational symmetry has been explicitly broken.  We take
initial conditions as before, except we introduce
different rescalings of the $x$, $y$ and $z$ coordinates in the
definition of $\bm{r}$.  As an additional numerical check, this run also
uses a smaller time step, $\Delta t = 0.05$.  Although the beat
pattern is slightly enhanced, likely indicating that we have started
further away from the true oscillon because of the nonspherical
deformation, we see that the system nonetheless converges to a very
similar configuration to the case without the rescaling.  Equivalent
behavior is seen when we make these two changes individually and when
we make other perturbations, such as variations of $\chi$.

Finally, we consider the topological properties of the electroweak
oscillon.  Unfortunately, as shown in \cite{Farhi}, there is no
unambiguous definition of the topological charge for solutions to the
equations of motion.  (Topological properties are typically studied
using vacuum-to-vacuum paths \cite{sphaleron}, which are clearly not
solutions to the equations of motion since they do not conserve energy.)
However, for any localized spatial configuration in which
the Higgs field never vanishes, the Higgs winding number is
unambiguously defined as
\begin{equation}
n = \frac{1}{24 \pi^2} \int \epsilon_{ijk} {\rm Tr\,} \left[
U^\dagger (\partial_i U) U^\dagger (\partial_j U) U^\dagger (\partial_k U)
\right]  d^3 x \,,
\end{equation}
where $U$ is the unique $SU(2)$ matrix associated with a nonvanishing
Higgs field $\Phi$, so that
\begin{equation}
\Phi = |\Phi| U \pmatrix{0\cr 1} \,.
\end{equation}
The Higgs winding number is a topological invariant, which can only
change with time if the Higgs field passes through zero at some point
in space.  The change in the Higgs winding is physically meaningful
and measures whether the fields have crossed the sphaleron barrier.
Because the electroweak oscillon contains only small-amplitude field
fluctuations, its Higgs winding is always zero and it does not
approach the sphaleron barrier.  Correspondingly, its topological
density
\begin{equation}
q=\frac{g^2}{64 \pi^2} \epsilon^{\mu \nu \lambda \sigma}
\bm{F}_{\mu \nu} \cdot \bm{F}_{\lambda \sigma}
\end{equation}
is small as well.  But the restriction to small amplitude does not
apply to its decays (induced, for example, by collision with another
oscillon), when the fields frequently exhibit an implosion to small
radii and large amplitudes before ultimately dispersing.  This
behavior is seen in Fig.\ \ref{fig:decay} for the oscillon's decay
when $\lambda$ is slightly less than one.  However, both this
particular decay and limited experiments with oscillon collisions have
not led to winding in the final Higgs field.  Current work continues
to investigate this possibility.

\section{Conclusions}

We have seen in detail the results of a numerical simulation
describing a long-lived, localized, oscillatory solution to the
equations of motion in the bosonic sector of the electroweak Standard
Model, for a Higgs mass that is twice the $W^\pm$ mass.  Compared to
the natural scales of the system, this solution has small field
amplitudes, large spatial extent, and large total energy.  In
the quantized theory, it would represent a coherent superposition of
many elementary particles, and thus is well described by the classical
analysis undertaken here.  Quantization of the small oscillations
around the classical solution would nonetheless be of interest,
as has been done for $Q$-ball oscillons in \cite{qqball}.  It would
also be desirable to incorporate fermion couplings, which have been
ignored here.  Such an analysis would require introducing chiral
fermions on the lattice, which is well known to be a difficult
problem, but one on which significant progress has been made in recent
years.  While one might expect the oscillon to be destabilized by
decay to light fermions, in the case of the photon coupling we have
seen that the analogous decay mechanism is highly suppressed.

Because it would require bringing many Higgs and gauge particles
together at once, forming such an oscillon would likely require large
energies available only in the early universe.  If extremely
long-lived, such an oscillon could be a dark matter or ultra-high
energy cosmic ray candidate.  A slow fermion decay mode would be of
interest for baryogenesis, since it could provide a mechanism for
fermions to be produced out of equilibrium, as is necessary to avoid
washout of particle/antiparticle asymmetry.  The oscillon has small
amplitude everywhere and thus remains far from the sphaleron
configuration, even though it has energy above the height of the
sphaleron barrier.  However, when induced to decay, for example by a
collision with another oscillon, the fields typically collapse to a
configuration with small radius and large energy density and field
amplitudes before dispersing.  Such decays could potentially cross the
sphaleron barrier and produce fermion number violation.  For
baryogenesis applications, one would also need to incorporate
interactions containing $C$ and $CP$ violation in the classical
effective action.

The spherical ansatz provided a crucial tool for obtaining 
the electroweak oscillon solution.  However, any search for oscillons
using a particular ansatz cannot guarantee that all solutions have
been found.  ``Emergent'' techniques, in which oscillons form
from generic initial conditions, offer the opportunity for more
comprehensive searches for oscillons, albeit at a higher computational
cost.  In simpler models, oscillons have been shown to emerge from phase
transitions \cite{Gleiserphase} and from thermal initial conditions in
an expanding universe \cite{emerge}.  Clearly, it would be desirable to
extend these techniques to the electroweak model.

The electroweak oscillon remains stable even when one would
expect it to decay, suggesting that there might exist other stable,
oscillatory solutions in the electroweak theory or its extensions,
either for generic or specific mass ratios.  While results for generic mass
ratios are clearly of broader applicability, a compelling result for
a specific mass ratio might suggest a preferred value of the Higgs mass.

\section{Acknowledgments}

It is a pleasure to thank E.\ Farhi, F.\ Ferrer, M.\ Gleiser,
A.\ Guth, R.\ R.\ Rosales, R.\ Stowell, J.\ Thorarinson,
and T.\ Vachaspati for helpful discussions, suggestions and
comments; P.\ Lubans, C.\ Rycroft, S.\ Sontum, and P.\ Weakleim for
Beowulf cluster technical assistance; and the Massachusetts Institute
of Technology (MIT) Center for Theoretical Physics for hospitality and
support while this work was being carried out.
N.\ G. was supported by National Science Foundation (NSF) grant
PHY-0555338, by a Cottrell College Science Award from Research
Corporation, and by Middlebury College.

Computational work was carried out on the Hewlett-Packard (HP) Opteron
cluster at the California NanoSystems Institute (CNSI) High
Performance Computing Facility at the University of California, Santa
Barbara (UCSB), supported by CNSI Computer Facilities and HP; the
Hoodoos cluster at Middlebury College; and the Applied Mathematics
Computational Lab cluster at MIT.  Access to the CNSI system was
made possible through the UCSB Kavli Institute for Theoretical Physics
Scholars Program, which is supported by NSF grant PHY99-07949.

\end{document}